# Computational Insight into the Complexation of DNA-Functionalized Gold Nanoparticles

J.Hingies Monisha [1)], V.Vasumathi[1)], Prabal K Maiti [2)]

[1)] PG & Research Department of Physics, Holy Cross College (Autonomous), Affiliated to Bharathidasan University, Tiruchirappalli-620002, Tamilnadu, India.

[2)] Center for Condensed Matter Theory, Department of Physics, Indian Institute of Science, Bangalore-560012, India.

**Abstract**

Ensuring the stability of the AuNP-gene complex until it reaches the target sites is a crucial factor for the success of gene therapy. Though different AuNP sizes and AuNP-to-DNA ratios are investigated for specific therapeutic needs, their role on the stability and packaging of AuNP-DNA complex remains unclear. In this study, we employ all-atom molecular dynamics simulations to investigate the influence of cationic ligand-functionalized AuNP (CAuNP) size and CAuNP-to-DNA ratio on DNA wrapping and binding affinity. The obtained results show that single DNA interacting with smaller CAuNPs exhibit greater bending and wrapping due to their higher curvature. However, when two DNAs bind to smaller CAuNPs, electrostatic repulsion prevents the effective wrapping which leads the DNAs to twist from their original orientation. Such behaviour is not observed with larger CAuNPs since their increased size may mitigates repulsive forces. Further, the analysis on axial bending angle reveals that smaller AuNPs induce sharper DNA bending and larger AuNPs promote smoother bending. In addition, the Potential of Mean Force (PMF) analysis confirms a stronger DNA binding affinity for larger AuNPs with affinity decreasing when two DNAs attach to a single CAuNP. Our results from the DNA loading capacity calculations provide insights into the maximum number of DNA molecules that can be loaded onto CAuNPs of a given size. These findings offer key insights into optimizing the size of AuNP and DNA-to-AuNP ratios for the development of efficient gene delivery systems.

## 1) Introduction

Gene therapy is the most promising medical approach that helps to treat and prevent various genetic diseases[1] and disorders[2-3], such as hemophilia[4-5], thalassemia[6], cystic fibrosis[7-8], immune deficiencies[9], Parkinson's disease[10], and cancer[11-12]. In this therapeutic approach, the mutated gene that is responsible for genetic defects is replaced with a healthy copy of the gene i.e., therapeutic gene[13-14]. Here, the most crucial step is delivering a therapeutic gene to the targeted body location where the mutated cells reside and it could be facilitated by various transfection vectors[15]. Generally, two kinds of vectors such viral and non-viral vectors have been employed for gene transfer, and both has its own advantage and disadvantages[16-17]. However, the non-viral vectors including polymers, lipids, and inorganic nanoparticles, have gained vast attention because of its properties like low toxicity, higher gene carrying capacity, low immune responses, and feasibility of large-scale production [18-19-20-17].

Especially gold nanoparticles (AuNP) stand out among other vectors due to their facile synthesis, precise control over size and structure, ease of surface functionalization, and excellent biocompatibility[21,22]. By tuning the size, shape, and functional ligands of AuNPs, one could control their bio-interactions and the corresponding properties, including cell internalization[23], excretion[24], cytotoxicity[25], and biodistribution[21,26]. These characteristics make AuNPs an optimal vector for gene therapy applications.

One of the crucial factors influencing the efficiency of gene therapy is the stability of the vector–gene complex during transport to the target cells [27]. Generally, gene stability issues like enzymatic degradation could be resolved by the compact wrapping i.e., compaction of genes around AuNP[28,29]. Studies have identified that compaction of genes on the AuNP, can be controlled by various factors, such as ionic strength of the solvent, charge and special arrangement of functionalising ligands on AuNP. For example, it has been demonstrated that DNA adopts a compact conformation around citrated-functionalized AuNPs in a higher ionic strength solvent [30]. Similarly, highly charged cationic ligand-functionalized AuNPs (CAuNPs) with a spherical ligand arrangement facilitate compact wrapping of DNA around the CAuNP surface [15].

Another critical factor influencing the stability of the Gene-AuNP complex is the binding strength between the gene and AuNP. Lower binding affinity may result in the premature release of DNA/RNA from AuNP before reaching the target cells, whereas higher binding strength could decelerate the DNA/RNA release rate, thereby impeding the efficiency of gene therapy[27,31,32]. As the attachment of gene to CAuNPs mainly driven by the electrostatic attraction, the binding affinity is directly correlated to the density of cationic ligands on the surface of the CAuNPs [33]. Moreover, studies have shown that the type of cationic head group of the ligands, the ionic strength of the solvent[34] and the temperature[35] can modulate the binding strength between genes and AuNPs[36].

Although several studies have identified the factors that influence the wrapping of DNA/RNA around CAuNPs and their binding affinity, the role of certain key factors in this behaviour remains unclear. Such key factors include AuNP size and the payload-to-carrier ratio, referring to the amount of DNA/RNA that can be efficiently loaded onto the AuNP surface. Generally, AuNPs of various sizes, ranging from 2 to 100 nm, are utilized in gene therapy[18]. These size variation lead to differences in surface curvature [37], which in turn impact DNA compaction, as DNA structure tends to adapt to the curvature of the AuNP during the binding. Additionally, it has been demonstrated that the arrangement of functionalizing ligands on AuNPs varies with their size[38]. Since the ligand arrangement on CAuNPs impacts the wrapping of DNA/RNA[15], it is crucial to study the role of AuNP size in gene wrapping and binding affinity. Furthermore, the carrier-to-gene ratio has been identified as a key determinant in the formation of carrier–gene complexes[39]. However, the precise impact of this ratio on the wrapping and binding affinity of these complexes has not yet been elucidated.

In addition, understanding how factors such as CAuNP size and the CAuNP-to-DNA ratio influence DNA wrapping and binding is critical for accurately determining DNA loading capacity. The extent of DNA wrapping and the strength of binding affinity directly impact the efficiency of DNA attachment to the CAuNP surface. For instance, incomplete DNA wrapping can result in unbound regions, while weak binding affinity may lead to unstable interactions. Both factors can significantly reduce the overall DNA loading capacity.

With this motivation, in this study, we have conducted all-atom molecular dynamics (MD) simulations to investigate the influence of AuNP size and the CAuNP-to-DNA ratio on the wrapping of DNA around CAuNPs and the binding strength between them. Specifically, we have considered small AuNPs with diameters of 2nm, 3nm, and 4nm, along with CAuNP-DNA complexes at two ratios, 1:1 and 1:2 (number of AuNP: number of DNA), for this study. To calculate the binding affinity between CAuNP and DNA, we have used the Umbrella sampling (US) simulations. The potential of mean force (PMF) obtained from these US simulations provide the free-energy barrier between the unbinding and binding states of CAuNP-DNA complex, which quantitatively reflects the binding affinity[40, 41]. The remainder of the paper is organized as follows: In the next section, we present the details of molecular system preparation and simulation procedures. In Section 3, we provide and discuss the results from our all-atom MD simulations. Finally, in Section 4, we summarize the findings and conclude with some future outlook.

## 2) Molecular Modelling Methodology

### 2.1. System preparation

The AuNPs of three different sizes: 2nm, 3nm and 4nm, were modelled using Vesta package[42]. We chose an 11-carbon alkanethiol with a terminal amine group (S-$(CH_2)_{11}$-$NH_3^+$) as a cationic ligand to functionalize AuNPs (henceforth referred as CAuNP). The ligand for CAuNPs was modelled using Amber's xleap program[43] and the partial atomic charges were assigned by first calculating the electrostatic potential (ESP) with Gaussian package [44], followed by deriving restrained electrostatic potential (RESP) charges using antechamber module[43] of AMBER. Each ligand was attributed a +1e charge and the total charges of functionalized AuNPs depend on the number of ligands attached to it. The AuNPs were functionalized with the appropriate number of ligands, such as 80, 175 and 370 for 2 nm, 3 nm, and 4 nm AuNPs, respectively, using Packmol package[45]. The number of ligands were chosen based on the previous adsorption and MD studies[46-47]. A dsDNA of 30 base pairs (TTCTACCAAAAGTGTATTTGGAAACTGCTC) was generated using Avogadro package[48]. This sequence is adapted from the crystal structure of DNA compacted around the nucleosome[15,49]. To avoid the influence of DNA flexibility on the wrapping and binding affinity, the DNA length was deliberately chosen to be shorter than its persistence length[15]. The dsDNA was positioned at a distance of approximately 15Å from the functionalized AuNP using the xleap module in Amber[43]. This configuration is referred to as the 1:1 complex, indicating that one CAuNP is interacting with one DNA molecule. To study the effect of DNA concentration (i.e., CAuNP-to-DNA ratio) on the compaction and binding affinity, 1:2 complexes were also generated for different sizes of AuNPs by placing second dsDNA at distance of 15Å of the CAuNP but diametrically opposite of the first dsDNA. In this 1:2 complex, the dsDNA on the left side of the CAuNP is referred to as L-DNA, and the right side one is referred as R-DNA. These CAuNP-DNA complexes were then solvated in the TIP3P [50] water box extending 20 Å from the solute in all the three directions. Subsequently, the appropriate number of Na$^+$ and Cl$^-$ ions were added to neutralize the system. The instantaneous snapshots of the initial configuration of all the systems, along with various nomenclature are shown in Fig. S1 of the supplementary information.

**2.2. Simulation Procedure**

All the molecular dynamics (MD) simulations were performed using the PMEMD module of Amber 20 software package[43]. The inter and intra-molecular interactions involving AuNPs and the functionalising ligands were described by the parameters from the Generalized amber force field (GAFF)[51], while for dsDNA we used OL15 force field[52-53]. Initially, each system was subjected to 1,000 steps of steepest descent minimization, followed by 1,000 steps of conjugate gradient minimization. During this energy minimization process, the CAuNP-DNA complexes were held fixed in their initial position using harmonic constraints with a force constant of 25 kcal/mol/Å to eliminate bad contacts between the water molecules themselves and with the DNA-CAuNP complex. Subsequently, the energy-minimized systems were slowly heated from 0 to 300 K over 100ps using a Berendsen thermostat[54] with a temperature coupling constant of 1ps. Following this, the equilibration step was performed in the NPT ensemble, applying a week harmonic constraint of 10 kcal/mol/Å$^2$ to the dsDNA to preserve its initial position. This step allows the ligands to self-assemble into a monolayer on the AuNP surface. After the equilibration, all constraints were released and the production run was carried out in the NPT ensemble for 100 ns using the Berendsen barostat with pressure coupling constant of 5.0 ps.

**2.3. PMF calculation using Umbrella sampling**

To determine the binding affinity of the CAuNP-DNA complex, we have calculated the PMF using the Umbrella sampling (US) simulation[55]. The US simulations were performed using PMEMD module of Amber package[43]. In the US technique, simulations are carried out using biased umbrella potential along the reaction coordinate (RC)[56]. The RC is segmented into overlapping windows, where simulation within each window improve sampling efficiency, allowing the precise calculation of the PMF[57]. The starting structures for the US simulation, the CAuNP-DNA complex were taken from the final configurations of equilibrium production run described above. During the US runs, the dsDNA was gradually pulled away from the CAuNP along the centre of mass (COM) distance in the X direction, until the dsDNA was completely separated from the CAuNP. For the dsDNA pulling simulation, we followed the protocol used in our earlier work.[57] The spacing between successive US windows were chosen as 0.5Å and the biasing harmonic potential of force constant value of 40 Kcal/mol imposed on each window. The US simulations were carried out in NVT ensemble for 2ns in each window, as this simulation length demonstrated to be sufficient in previously reported studies [58-59].

**2.4. System analysis**

All the simulated systems were analysed using Amber's CPPTRAJ program[60] and in-house scripts. The structural analysis of the dsDNA base pair parameters was performed using Curves+ package[61]. The weighted histogram analysis method (WHAM)[62] was used to postprocess the umbrella sampling to calculate the PMF along the reaction coordinate. Visual Molecular Dynamics program (VMD)[63] was used to generate all the snapshots of the simulated system.

**3) Results and Discussion**

**3.1) Microscopic picture of the complex**

In order to understand how dsDNA (both one and two dsDNAs) wraps around the CAuNPs of various sizes we have presented the equilibrated configuration of various systems in Fig. 1. From the figures, we clearly observe the variation in the degree of wrapping of dsDNA with respect to AuNP size and dsDNA concentration. In case of 1:1 complex, dsDNA exhibits complete wrapping around CAuNPs of all the sizes. In contrast, in the case of 1:2 complex, the dsDNA exhibits incomplete wrapping, this effect being particularly pronounced for smaller CAuNPs. For example, in the case of CAuNP_2nm, the top several base pairs (nearly 6 base pairs) of L-DNA and some bottom base pairs of R-DNA remains unbounded from the CAuNP surface. A similar partial wrapping is observed in the case of CAuNP_3nm, where few base pairs in the bottom region of L-DNA do not interact with CAuNP surface. In the case of CAuNP_4nm also, dsDNA exhibits incomplete wrapping, but the length of the unbound regions are minimal, with very few base pairs at the top of R-DNA not in contact with the CAuNP_4nm surface.

Interestingly, we have also observed the twisting of dsDNAs from their initial position when they interact with smaller AuNPs, i.e., the CAuNP_2nm and CAuNP_3nm, in a 1:2 complex. However, such a behaviour is absent in the case of the larger CAuNP_4nm. When two dsDNAs come in close contact with the CAuNP, both strive to wrap around the nanoparticle in a compact manner. When two dsDNAs interact with CAuNP_2nm and CAuNP_3nm, due to the smaller size of CAuNPs, the top and bottom edges of the dsDNAs are extend beyond the CAuNP surface and are freely exposed to one another(see Fig. S1 d-e and Fig. 2 a). Thus, the negatively charged phosphate atoms on the exposed top and bottom edges of the both the dsDNAs would repel each other. Despite this repulsion, the attraction of their centre region to CAuNPs (i.e., the electrostatic attraction between nitrogen atoms of CAuNPs and P atoms of dsDNA) enables the DNAs to twist from their oriented (i.e., vertical) axis to wrap around the smaller CAuNPs (see Fig. 1d and 1e).

However, in the case of CAuNP_4nm, the length of dsDNA extending beyond the CAuNP_4nm surface (i.e., the freely exposed top and bottom edge regions of DNAs) appears to be very minimal. Also, the size of CAuNP_4nm is large enough to hinder the repulsion between both the DNAs (see Fig S1 f and Fig. 2 b). Therefore, when both the DNAs approach the CAuNP_4nm, they retain their initial vertical orientation without twisting upon attachment (see Fig. 1f).

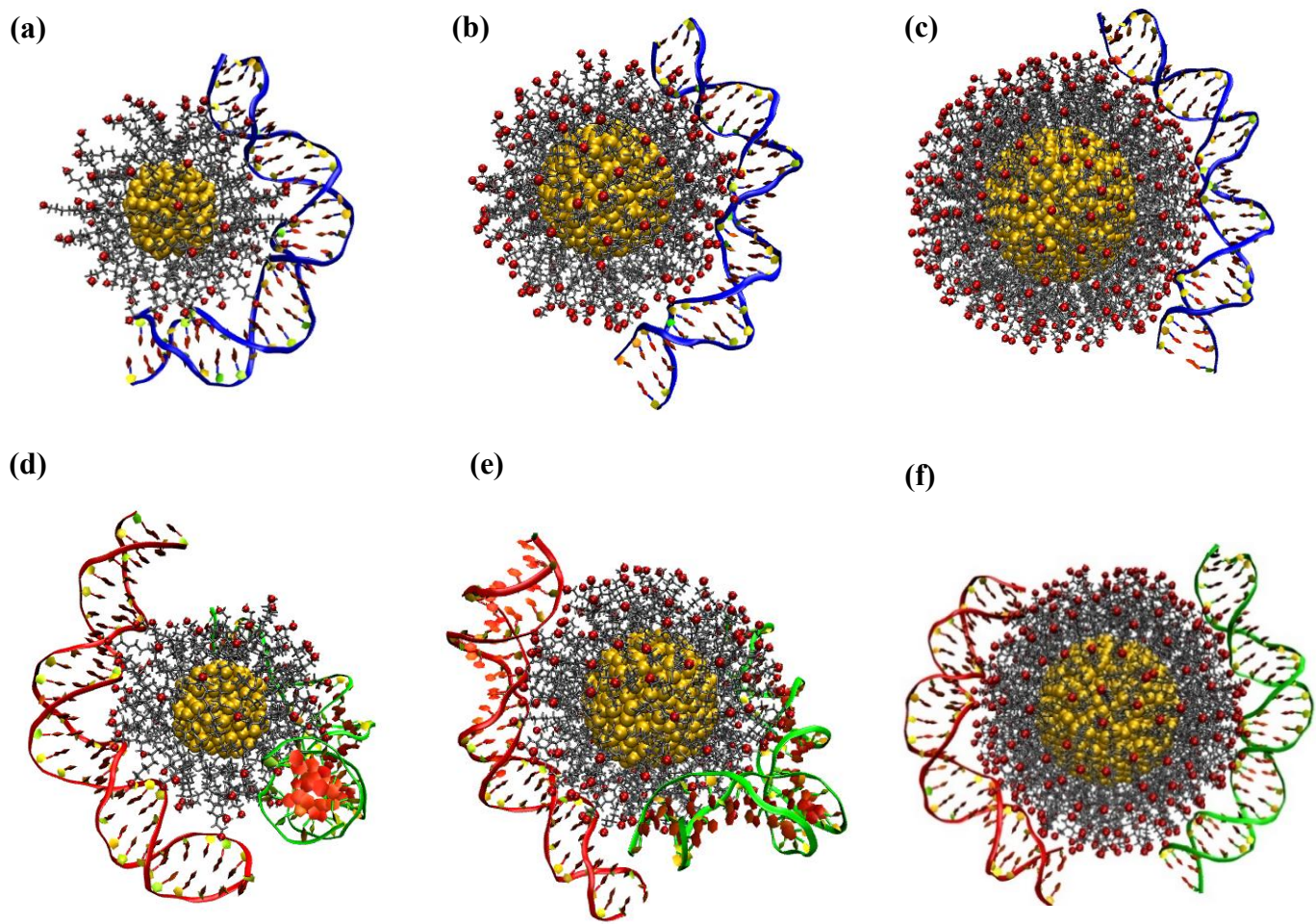

**Fig. 1.** Snapshots of the equilibrated configurations of all the cases (a) CAuNP_2nm, (b) CAuNP_3nm , and (c) CAuNP_4nm in the 1:1 complex. (d) CAuNP_2nm, (e) CAuNP_3nm, and (f) CAuNP_4nm in the 1:2 complex. In a 1:2 complex, green-colored dsDNA is referred to as R-DNA (right-side DNA), and red-colored DNA as L-DNA (left-side dsDNA).

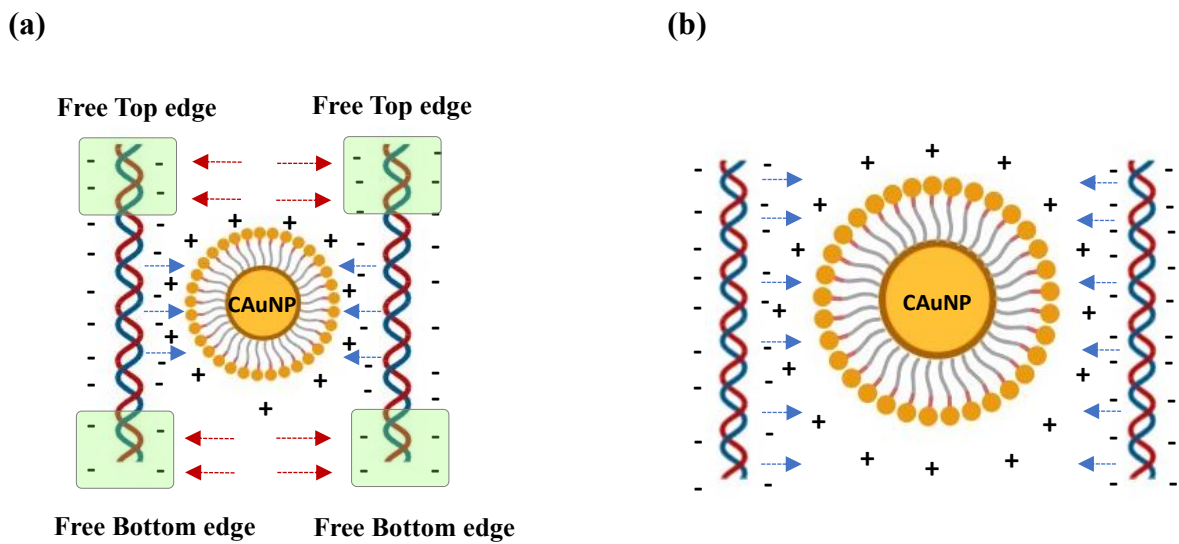

**Fig. 2.** A Two-dimensional schematic illustration depicting the interaction between dsDNAs and CAuNPs during binding. (a) dsDNAs with free top and bottom edges interacting with a small CAuNP, illustrating the repulsion between the terminal ends and the attraction of the dsDNA central region toward CAuNP and, (b) dsDNAs without free edges interacting with a larger CAuNP, showing uniform attraction of the dsDNA toward the CAuNP surface.

### 3.2) Number of close contacts between CAuNP and dsDNA

As the CAuNPs are functionalized with ligands having cationic end groups, their interactions with dsDNA are mainly driven by electrostatic attraction between the positively charged ammonium end groups and the negatively charged phosphate (P) backbone of the dsDNA[33,64]. The number of close contacts between the cationic terminal group of CAuNPs and the anionic P atoms of dsDNA directly characterizes the wrapping and binding of dsDNA around CAuNP. Thus, the average number of terminal Nitrogen (N) atoms of CAuNP within 5Å distance from the P atoms of dsDNA are calculated for all the systems and the results are presented in the Table 1. For both the 1:1 and 1:2 complexes, the number of close contacts ($N_c$) between CAuNP and dsDNA increases with the size of CAuNPs. Generally, the number of ligands attached to the CAuNPs, and consequently the number of cationic N atoms, tends to increase as the size of the AuNP increases. Also, for the larger CAuNPs, the spacing between terminal groups of the attached ligands is shorter compared to that of smaller CAuNPs[38]. To validate this, we calculated the radial distribution function (RDF) between the N-N atoms of the CAuNP, and the results are illustrated in Fig. 3. From the analysis, the highly probable distance at which another N atom is located in its immediate vicinity of each N atom for CAuNP_2nm, CAuNP_3nm, and CAuNP_4nm is found to be 7 Å, 5.9 Å and 5.4 Å, respectively (see Fig. 3 b). The reduced spacing between N atoms for larger AuNPs leads to a denser and more uniform coverage of cationic groups on larger AuNPs. As a result, each P atom in the dsDNA can interact with a greater number of terminal N atoms in the larger CAuNPs.

This is further illustrated by the density plot showing the distribution of N atoms located within 5 Å distance from the P atoms in the dsDNA, as shown in Fig. S2. The density plot clearly shows that compared to smaller CAuNPs, larger CAuNPs have relatively more cationic amines positioned near each P atoms of the dsDNA. Thus, dsDNA binding increases with CAuNP size, as more N atoms interact with the P atoms of the dsDNA in larger CAuNPs.

Further, we studied the effect of dsDNA ratio. Compared to the 1:1 complex, the dsDNAs in the 1:2 complex exhibits lower number of close contacts for all three different sizes of CAuNP. This effect is particularly significant for smaller CAuNPs, such as CAuNP_2nm and CAuNP_3nm. Also, the two dsDNAs in the 1:2 complex that interact with these smaller CAuNPs display different number of close contacts. The observed lower number of close contacts is due to the partial wrapping of dsDNAs onto the CAuNP, which is evident from the instantaneous snapshots of the equilibrated configurations of 1:2 complexes shown in Fig. 1 (d-f). As shown in the figure, the top and bottom ends of dsDNAs in 1:2 complex do not come in close contact to the surface of smaller CAuNPs. This observation can be further supported by measuring the number of N atoms in close contact with the P atoms at the top and bottom edges of the dsDNA. For this calculation, we have considered the 5-base pair above the first turn of dsDNA as a top edge and the 5-base pairs below the last turn of dsDNA as a bottom edge. The calculated values are presented in Table 2, where $N_{C\_T}$ and $N_{C\_B}$ denote the number of N atoms in close contact with the top and bottom ends of the dsDNA, respectively. For all the different sizes of CAuNPs, the $N_{C\_T}$ and the $N_{C\_B}$ values are lower for the 1:2 complex compared to the values for the 1:1 complex. These results suggest that the terminal segments of the dsDNAs in the 1:2 complex show a lower number of contacts to the surface of the CAuNP compared to the 1:1 complex, with several segments remaining partially or fully unbound.

However, for the CAuNP_4nm system, only a minimal difference is observed in the values of $N_C$, $N_{C\_T}$ and $N_{C\_B}$ between the 1:1 and 1:2 complexes, indicating that both dsDNAs exhibit similarly strong attachment to the CAuNP_4nm surface, closely resembling the wrapping behaviour observed in the 1:1 complex.

**Table 1.** The average number of terminal nitrogen (N) atoms of CAuNP within 5Å distance from the phosphate atoms (P) of dsDNA, where dsDNAs in 1:1 and 1:2 complex are referred to as DNA 1:1 and DNA 1:2 (R-DNA1:2 and L-DNA1:2).

|  | Number of N atoms of AuNP within 5Å distance from the P atoms of DNA ($N_c$) | | |
| --- | --- | --- | --- |
|  | CAuNP_2nm | CAuNP_3nm | CAuNP_4nm |
| **DNA 1:1** | 19.99 ± 2.28 | 27.15 ± 2.00 | 30.27 ± 2.48 |
| **R-DNA 1:2** | 18.62 ± 2.35 | 23.89 ± 2.35 | 29.56 ± 1.60 |
| **L-DNA 1:2** | 15.66 ± 1.99 | 21.36 ± 1.71 | 28.50 ± 2.21 |

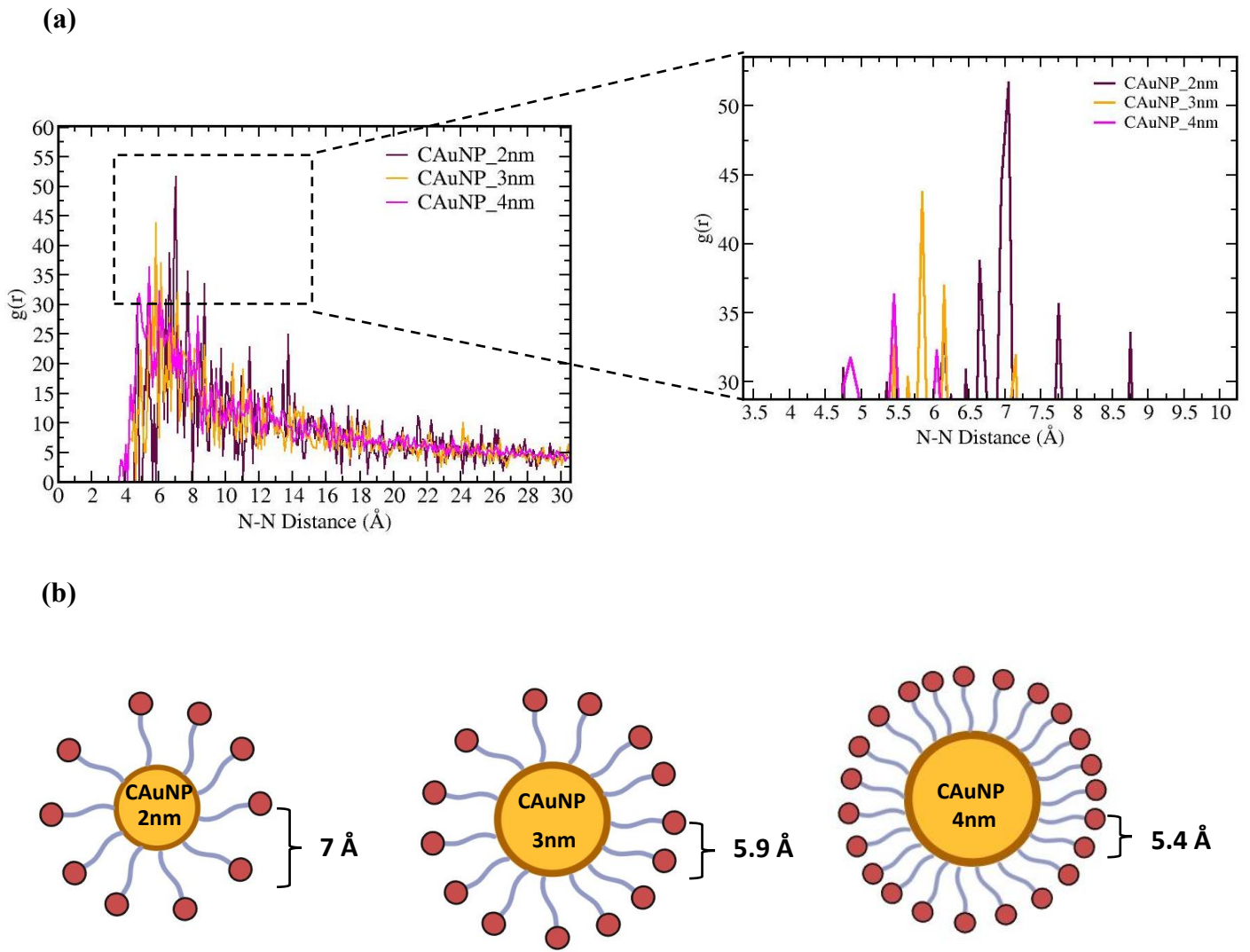

**Fig 3.** (a) Radial Pair Distribution (RDF) between the terminal N atoms of CAuNPs. (b) Two-dimensional schematic representation of the distances between terminal Nitrogen atoms in the CAuNP_2nm, CAuNP_3nm and C_AuNP_4nm.

**Table 2.** The average number of terminal nitrogen atoms of CAuNP within 5Å distance from the phosphate atoms of top and bottom edges in the dsDNA. Where dsDNAs in 1:1 and 1:2 complex are referred to as DNA 1:1 and DNA 1:2 (R-DNA1:2 and L-DNA1:2), respectively.

|  | Number of N atoms of AuNP within 5Å distance from the P atoms of top and bottom edges in the DNA ||||||
|---|---|---|---|---|---|---|
|  | DNA 1:1 || R-DNA 1:2 || L-DNA 1:2 ||
|  | Top ($N_{C\_T}$) | Bottom ($N_{C\_B}$) | Top ($N_{C\_T}$) | Bottom ($N_{C\_B}$) | Top ($N_{C\_T}$) | Bottom ($N_{C\_B}$) |
| CAuNP_2nm | 4.96 ±2.14 | 6.75± 1.68 | 4.22 ± 1.48 | 2.3±1.58 | 0 | 1.76 ± 2.88 |
| CAuNP_3nm | 6.73±1.34 | 6.48±2.08 | 4.76 ± 1.94 | 5.77± 1.86 | 3.7±2.36 | 2.14± 1.84 |
| CAuNP_4nm | 6.77± 1.88 | 7.88± 1.46 | 5.12± 1.64 | 6.64± 1.24 | 5.92± 2.82 | 5.82±1.98 |

### 3.3) Bending of dsDNA around CAuNP

### 3.3.1) Global bending

The compact wrapping of dsDNA involves its bending[65] and condensation into a smaller spatial volume[66, 67, 68, 69] If dsDNA bends more while attached to CAuNP, it would be compactly wrapped into a smaller volume. Hence, to quantify the compactness of dsDNA wrapping, we calculated its global bending angle, defined as the total degree of bending along the DNA helical axis. This is determined by adding the bending angles between the adjacent base pairs using the Curves+ program[61]. The bar graph in Fig.4 represents the average global bending of dsDNAs in the 1:1 and 1:2 complex for different sizes of CAuNPs. As shown in figure, the bending of dsDNA in the 1:1 complex decreases with increasing the size of CAuNPs. This is due to the inverse correlation between the CAuNPs size and its curvature i.e. the smaller CAuNPs have a higher curvature than the larger CAuNPs. When the dsDNA approaches the CAuNP, its structure aligns with the curvature of CAuNPs, as a result the dsDNA bends more and exhibits the compact wrapping around the smaller CAuNP.

This trend is not visible in the case of 1:2 complex, where the dsDNAs are interacting with smaller CAuNPs i.e., CAuNP_2nm shows a lesser bending compared to the CAuNP_3nm and CAuNP_4nm . This behaviour is due to the incomplete wrapping of dsDNAs around the CAuNP_2nm. Although the CAuNP_2nm possess higher curvature, the top and bottom ends of DNAs remain unbound from the CAuNP surface. As a result, both the dsDNA cannot achieve compactly wrapping conformation on the CAuNP_2nm surface. Here, the dsDNAs that interact with CAuNP_3nm also exhibit incomplete wrapping; however, this effect is less pronounced compared to that observed with the CAuNP_2nm. Hence, the dsDNAs are better able to follow the curvature of CAuNP_3nm, leading to a relatively more compact wrapping than in the CAuNP_2nm case. While in the CAuNP_4nm, both the dsDNAs are almost fully attached to the CAuNP surface, which enables it to closely follow the curvature. Therefore, in the cases of CAuNP_3nm and CAuNP_4nm, the inverse correlation between CAuNP size and dsDNA bending is preserved, such that increasing CAuNP size results in decreased DNA bending.

Moreover, the dsDNA in the 1:1 complex exhibits a higher degree of bending compared to the 1:2 complex, irrespective of CAuNP size. The reason for the observed reduction in the DNA bending is the electrostatic repulsion between the dsDNAs. Studies have shown that when the two or more dsDNAs interact with the same nanoparticle, they repel each other due to their negative charges[70]. This repulsion leads to the reduction in the association of the dsDNAs with CAuNPs, resulting in the reduced bending of both the dsDNAs.

Although the wrapping behaviour varies with CAuNP size—showing incomplete DNA wrapping for smaller CAuNPs and complete wrapping for larger ones—one of the dsDNAs in the 1:2 complex consistently exhibits greater bending than the other, regardless of CAuNP size. Specifically, the dsDNA residing on the left side of the CAuNPs in the 1:2 complex, i.e., R-DNA1:2, bends more than the dsDNAs on the right side, i.e., L-DNA1:2. This can be attributed to the number of major and minor grooves on the dsDNA side that are exposed to CAuNP surface. In the minor grooves, the anionic P atoms of dsDNA are closer to the cationic amines of the CAuNP, as these grooves are less deep and less wide than the major grooves. In the 1:2 complex, the exposed facet of R-DNA1:2 to the CAuNP has a greater number of minor grooves, whereas in the case of L-DNA1:2, major grooves are dominant (see Fig. S3). This facilitates the higher bending of R-DNA1:2 in the 1:2 complex.

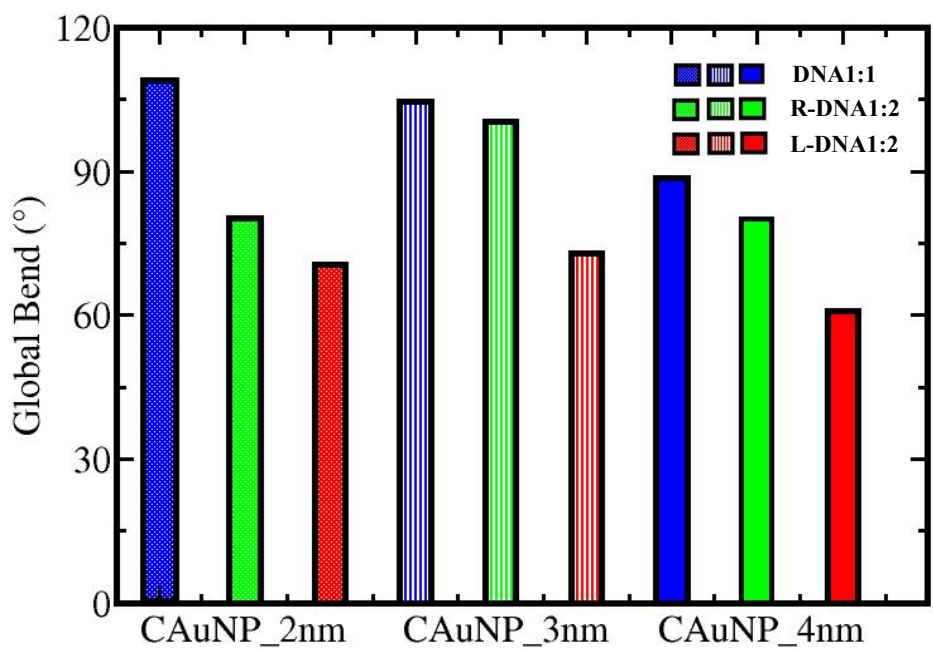

**Fig. 4.** Global bending angle of dsDNA of all the system simulated in this work.

**3.3.2) Axial bending**

Studies shows that the interaction of DNAs with DNA binding protein often induces two types of bending: smooth and sharp bending[71]. For example, DNA binding to the histone octamer[72, 73], exhibits smooth bending[15]. In contrast, DNA binding to integration host factor (IHF) proteins results in sharp bending[74]. In some cases sharp bending can cause the disruption of base pairs, which leads to the formation of kinks in the DNA[75]. These kinks can make DNA less susceptible to the damage[76]. Thus, to clearly understand the local conformational changes in the dsDNA, we have measured the angle between the adjacent base pairs along the helical axis, i.e., the axial bend angle, which is shown in Fig. 5. Except for the dsDNA in the 1:1 complex for the CAuNP_2nm, in all other cases, the dsDNAs show a uniform distribution of bending along the helical axis. Also, the range of axial bend angles of these DNAs is comparable to the axial bending angles of DNA bound to the histone octamer, as reported in a previous MD study conducted by Nash et. al.[15]. Hence, one can say that CAuNP_3nm and CAuNP_4nm induce smooth bending of dsDNA due to their relatively lower surface curvature. In contrast, the higher curvature of CAuNP_2nm forces the DNA in 1:1 complex to conform more tightly to the CAuNP surface, resulting in sharp bending. However, in the case of the 1:2 complex, CAuNP_2nm is unable to induce sharp bending in either DNA due to electrostatic repulsion between them.

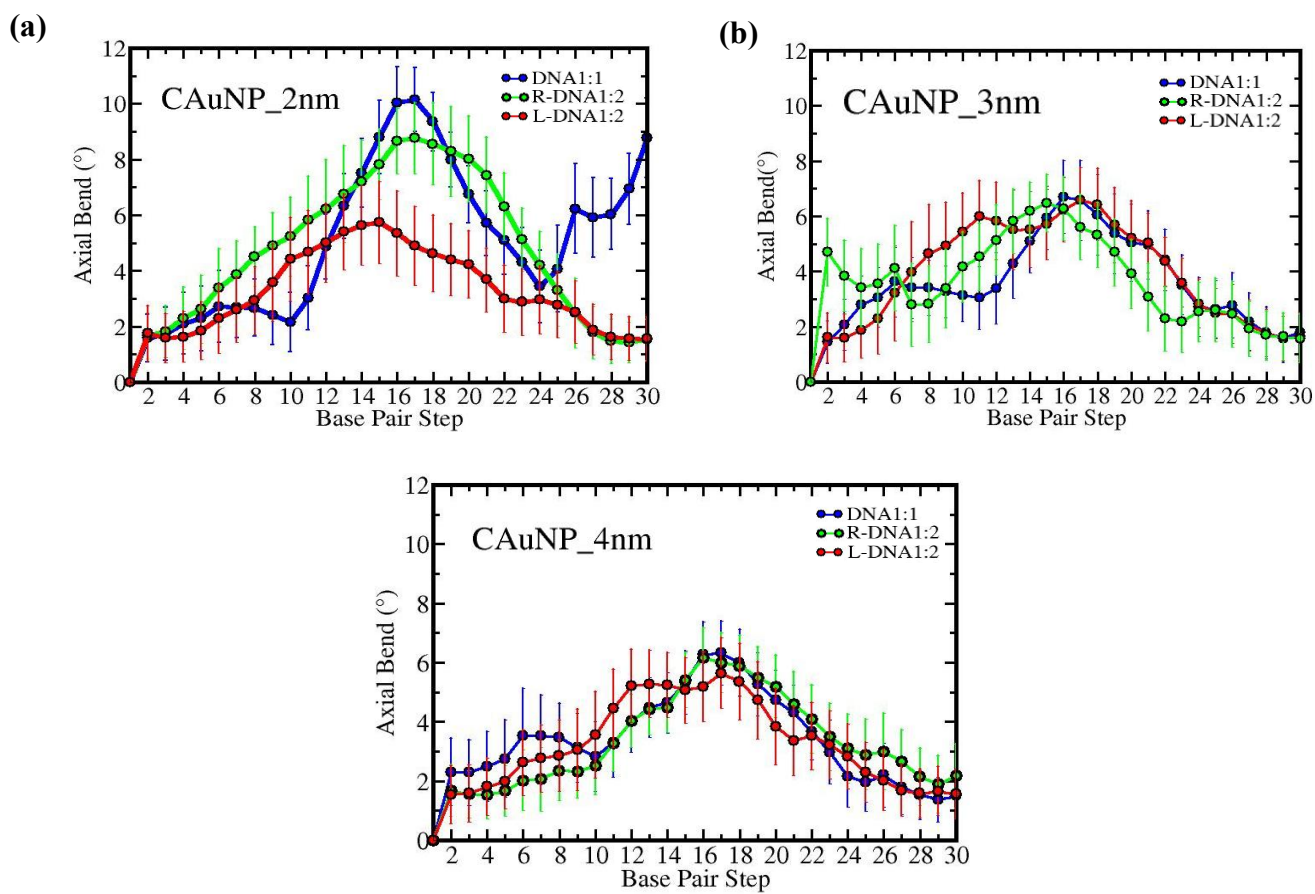

**Fig. 5.** Axial bending angles of dsDNA for 1:1 and 1:2 complex when bound to (a) CAuNP_2nm, (b) CAuNP_3nm , and (c) CAuNP_4nm.

## 3.4) Number of dsDNA Hydrogen bonds

To check whether the dsDNA bending upon binding with CAuNPs influences its structural stability, we have calculated the number of hydrogen bonds ($N_H$) between the complementary bases on the two strands of the dsDNA (see Table 3). Since hydrogen bonds hold two strands together, the overall structural integrity of dsDNA depends on them, and any considerable changes in the hydrogen bond could have a profound impact on dsDNA stability[77]. We have also calculated the $N_H$ value of canonical dsDNA simulated in a water box with an appropriate number of ions to serve as a reference value (see Fig. S4). This reference value is used to compare with the $N_H$ values of dsDNA interacting with CAuNP to evaluate the structural instability induced by bending. The calculated number of hydrogen bonds are presented in Table 3, which shows that the $N_H$ value in the case of CAuNP_4nm is very similar to the reference value, while a decrease in the $N_H$ count is observed with decreasing the size of the CAuNP. Notably, the CAuNP_2nm system in the 1:1 configuration exhibits the slightly low $N_H$ value, which is attributed to its sharp bending behaviour. Since the deviation from the reference $N_H$ value is minimal, reformation of hydrogen bonds can occur once the dsDNA unbinds from the CAuNP. To verify this, we measured the number of hydrogen bonds in the dsDNA that had detached from the CAuNP_2nm during the US simulation and found it to be 68.12, which is nearly identical to the reference value. This observation indicates that the bending of DNA during interaction with CAuNPs of different sizes does not significantly affect its structural stability, even in cases of sharp bending.

The number of hydrogen bonds in dsDNA, which is simulated in bulk with appropriate number of counter ions, is 68.68 ± 2.52.

**Table 3.** Number of hydrogen bonds between the complementary bases of the two strands of the dsDNA.

|  | Number of hydrogen bonds in dsDNA | | |
| --- | --- | --- | --- |
|  | CAuNP_2nm | CAuNP_3nm | CAuNP_4nm |
| **DNA1:1** | 65.34 ± 2.35 | 66.22 ± 2.31 | 68.36 ± 2.20 |
| **R-DNA1:2** | 66.42 ± 2.37 | 67.71 ± 2.25 | 68.48 ± 2.15 |
| **L-DNA1:2** | 66.22 ± 2.28 | 67.24 ± 2.39 | 68.28 ± 2.37 |

### 3.5) Binding affinity using Potential of mean force (PMF)

To determine the binding strength between dsDNA and CAuNP for all the systems simulated in this work, we performed PMF calculation using the US technique. The calculated PMF profiles for all the systems as a function of the reaction coordinate are shown in Fig. 6. The binding free energy is measured from the PMF plot by taking the difference between the largest and smallest PMF values corresponding to the unbound and the bound state of the dsDNA in the CAuNP-DNA complexes (see Fig S5), which is also provided in Table 4.

As depicted in Table 4, for both 1:1 and 1:2 complex, the binding free energy increased with increase in size of CAuNPs: CAuNP_4nm exhibits higher binding energy. The CAuNP_4nm possesses a greater surface area, allowing for a higher density of cationic end groups closely packed on their surface. Consequently, a greater number of cationic end groups interact with each negatively charged P atoms of the dsDNA (see Fig S2 c), which significantly enhances the overall binding affinity between CAuNP_4nm and dsDNA. In contrast, when the size of the AuNPs decreases, both the available surface area and the packing density of cationic end groups are reduced. This leads to a comparatively lower number of cationic end groups available for interaction with the P atoms of dsDNA (see Fig S2 a-b). As a result, CAuNP_2nm and CAuNP_3nm exhibit reduced binding affinities compared to CAuNP_4nm.

Furthermore, for all the CAuNP sizes, the dsDNAs in the 1:2 complex show lower binding free energy than the dsDNA in the 1:1 complex. Here, it should be noted that in the 1:2 complex, only R-DNA is pulled, while L-DNA remains bound to the CAuNP during the US simulation. Thus, from the results it is apparent that presence of one DNA affects the binding affinity of another DNA. In essence, the binding affinity of dsDNA on CAuNPs gets weaker, when more than one dsDNA interacts with CAuNP at the same time. This could be due to the electrostatic repulsion between the dsDNAs in the 1:2 complex.

Moreover, the disparity in binding free energies between the 1:1 and 1:2 complexes decrease as the size of the CAuNP increases. This difference is particularly pronounced for the CAuNP_2nm. Because, as smaller size CAuNPs couldn't completely screen one dsDNA from another, both dsDNAs repel each other, which results in the significant reduction in binding affinity of the second dsDNA in the 1:2 complex compared to the single dsDNA in the 1:1 complex. While for the CAuNP_3nm and CAuNP_4nm systems, due to their increased size, screening of DNAs gets increased, which leads to the lowering of repulsion between the DNAs resulting the equal binding affinity for second DNA in 1:2 complex and single DNA in 1:1 complex.

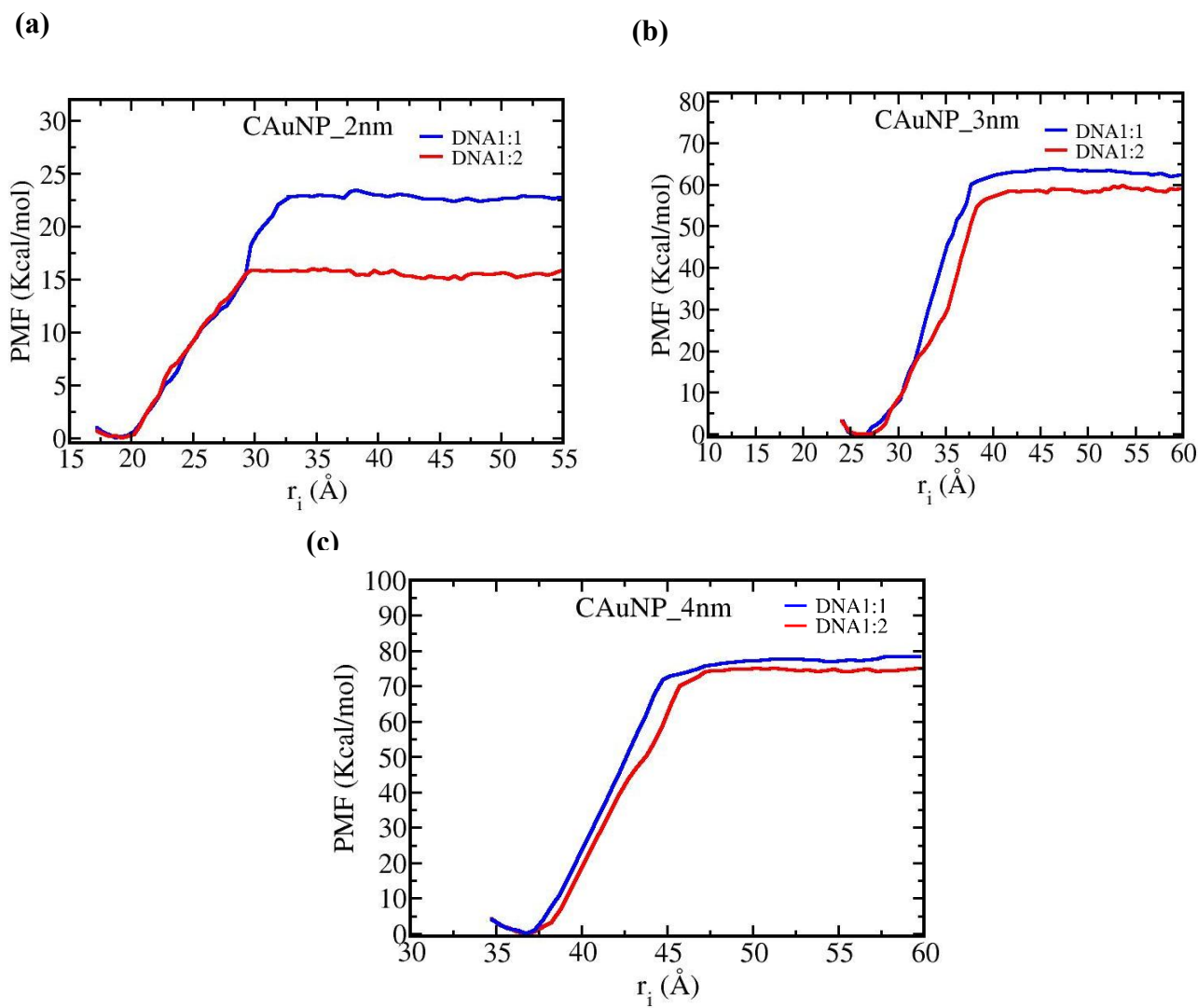

**Fig. 6.** PMF curve obtained from the US simulation for (a) CAuNP_2nm, (b) CAuNP_3nm and (c) CAuNP_4nm.

**Table 4.** Binding Free energy values obtained from US simulation.

|  | Binding Free energy ΔG (Kcal/mol) | | |
| --- | --- | --- | --- |
|  | **CAuNP_2nm** | **CAuNP_3nm** | **CAuNP_4nm** |
| **DNA 1:1** | -22.80 | -62.95 | -78.23 |
| **DNA 1:2** | -15.82 | -58.67 | -74.76 |

### 3.6) Percentage of DNA coverage on the circumference of CAuNPs

To gain a deeper understanding of why twisting of dsDNAs occurs with smaller CAuNPs but not with larger ones in a 1:2 complex, we measured the percentage of single DNA coverage on the circumference of CAuNPs. In this approach, we assumed the CAuNP to be circular and calculated its outer circular length, i.e., circumference, from its radius ($C = 2\pi r$).

The radius of CAuNP can be estimated by calculating the distance between the center of mass of the AuNP and the nitrogen atoms of each cationic ligand. The probability distribution of these distances is obtained for all CAuNP sizes (see Fig S6), and the distance corresponding to the maximum probability is taken as the radius, as it represents the CAuNP surface.

By relating the length of the dsDNA to the circumference of CAuNPs, we can determine the space occupied by the dsDNA i.e., the DNA's length coverage on CAuNP's circumference. The percentage of single DNA length coverage on the circumference of the CAuNPs is calculated using the below formula

$$\% \text{ of DNA coverage on the circumference of CAuNP} = \frac{L_{DNA}}{C_{CAuNP}} \times 100 \quad (1)$$

Where, $L_{DNA}$ and $C_{CAuNP}$ are the length of the dsDNA and the circumference of CAuNP, respectively.

The calculated radius, circumference of CAuNP of all the sizes, length of the dsDNA and single DNA's length coverage on different sizes of CAuNPs, are summarized in Table 5. As seen in table 5, the single DNA coverage on the circumference of the CAuNP is decreased with the size increment of CAuNPs. For CAuNP_2nm and CAuNP_3nm systems the percentage of single DNA's length coverage is more than 50%, indicating that only one dsDNA can be optimally wrapped around its circumference (see Fig. 7 a).

Since a single dsDNA is capable of occupying more than half of the circumference of a smaller CAuNP, the remaining circumference becomes insufficient to accommodate a second dsDNA in a similar orientation. When two dsDNAs attempt to bind simultaneously to the smaller CAuNP surface, the limited circumference prevents both from attaching vertically. Also, the excess free top and bottom ends of the both the dsDNAs experience electrostatic repulsion. This spatial constraint and the free end repulsion induce one of the dsDNAs to twist from its initial vertical orientation and wrap around the horizontal circumference of the smaller CAuNPs. Whereas in the case of larger CAuNP_4nm, the percentage of single DNA's length coverage is lesser than 50% (see Fig. 7 b), which depicts that two dsDNAs can be sufficiently attached to the surface of CAuNP without any twisting of DNAs from their original orientation.

**Table 5.** Radius (r) and circumference of CAuNP of various sizes ($C_{CAuNP}$), DNA length ($L_{DNA}$), and calculated single DNA coverage on the circumference of different sizes of CAuNPs.

|  | r (Å) | $C_{CAuNP}$ (Å) | $L_{DNA}$ (Å) | Percentage of single DNA length coverage (%) |
|---|---|---|---|---|
| CAuNP_2nm | 22.95 | 144.19 | 97.33 | 67.50 |
| CAuNP_3nm | 29.37 | 184.53 | 97.53 | 52.85 |
| CAuNP_4nm | 36.95 | 232.16 | 97.49 | 41.99 |

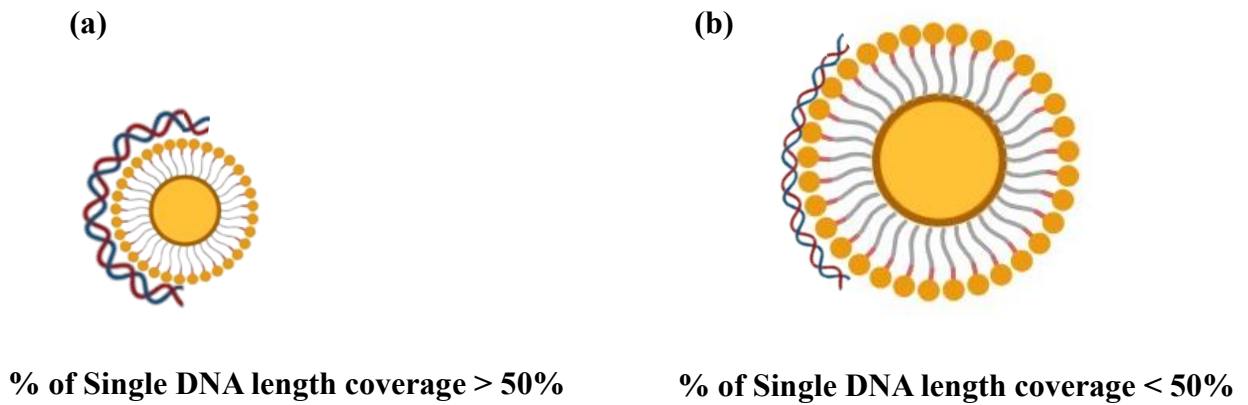

(a) % of Single DNA length coverage > 50%   (b) % of Single DNA length coverage < 50%

**Fig. 7.** Schematic illustration of single DNA coverage (a) greater than 50%, and (b) less than 50%.

### 3.7) Loading capacity

To estimate the DNA loading capacity of CAuNPs, we relate the available total CAuNP surface area to the surface area occupied by a single dsDNA molecule. Based on this relationship, we can predict the number of dsDNAs that can bind to CAuNPs of various sizes. Thus, we evaluated the surface area of CAuNPs and the surface coverage of a single DNA on CAuNPs in the 1:1 complex across all CAuNP sizes, and subsequently calculated the DNA loading capacity using the following formula:

$$N_{DNA} = \frac{S_{CAuNP}}{S_{DNA\_rep}} \qquad (2)$$

Where, $N_{DNA}$, $S_{CAuNP}$, and $S_{DNA\_rep}$, are the loading capacity of CAuNPs (i.e., number of DNAs that can be loaded to CAuNPs), surface area of the CAuNPs, and surface area of the CAuNP occupied by the DNA along with its repulsive zone, respectively.

In the denominator of Equation 2, instead of considering only the surface coverage of DNA on CAuNP ($S_{DNA}$), we account for the combined area occupied by the DNA and its corresponding repulsive zone ($S_{DNA\_rep}$). Because experimental and MD studies have reported that when two dsDNAs approach each other, they maintain a certain distance (inter-DNA spacing)[78,79], ranging from 10 to 15Å[80], due to the repulsion between them. Hence, while calculating the dsDNA coverage on the CAuNPs surface (i.e., $S_{DNA}$), it is essential to account for the repulsive zone associated with a dsDNA.

We have measured the surface area of the CAuNPs and the single DNA coverage on the surface of CAuNP by performing the Solvent Accessible Surface Area (SASA) calculation. However, the surface area of CAuNP covered by the repulsive zone (extending both longitudinally and transversely around the dsDNA; see Fig. S7) cannot be directly calculated from the SASA calculation. Hence, it is estimated by calculating the percentage of repulsive area in actual area of dsDNA and then multiplying the obtained percentage with calculated SASA value of dsDNA coverage on CAuNP surface. For that, the area of the dsDNA ($A_{DNA}$) was estimated by approximating dsDNA as a two-dimensional rectangle as follows.

$$A_{DNA} = L_{DNA} \times W_{DNA} \qquad (3)$$

Where $A_{DNA}$, $L_{DNA}$ and $W_{DNA}$ are the surface area, length and width of the dsDNA, respectively.

Then, the same rectangular approximation was used to estimate the total area encompassing actual dsDNA and its surrounding repulsive zone ($A_{DNA\_rep}$) as follows.

$$A_{DNA\_rep} = (L_{DNA} + D_{rep}) \times (W_{DNA} + D_{rep}) \qquad (4)$$

Where $A_{DNA\_rep}$ and $D_{rep}$ denote the area of dsDNA along with the repulsive zone and inter DNA spacing, respectively.

Then the area of the repulsive zone alone ($A_{rep}$) was obtained by subtracting the dsDNA area from the total area.

$$A_{rep} = A_{DNA\_rep} - A_{DNA} \qquad (5)$$

Using the actual dsDNA's length as 97 Å, width as 20 Å, and the repulsive distance as 12 Å (a rounded-down value from the reported 10–15 Å inter-DNA spacing) in the above equations, the percentage of repulsive zone is calculated to be ~80%. The step-by-step calculation of the repulsive zone percentage are provided in the Supporting Information Section S1.

Now, the surface area of the CAuNP covered by the dsDNA's repulsive zone ($S_{rep}$) is calculated by multiplying the SASA-derived value of the CAuNP surface area occupied by the single dsDNA ($S_{DNA}$) with the percentage of repulsive zone. Finally, the surface area of CAuNP occupied by dsDNA and its corresponding repulsive zone ($S_{DNA\_rep}$) is calculated by summing $S_{DNA}$ and $S_{rep}$.

The calculated $S_{CAuNP}$, $S_{DNA}$, $S_{rep}$, $S_{DNA\_rep}$, and $N_{DNA}$ values are given in Table 6. As presented in the table, the DNA loading capacity of CAuNPs increases with their size, which is attributed to the corresponding increase in the available surface area for DNA attachment. Here, the loading capacity of CAuNPs was calculated from the surface area information of both CAuNPs and DNA, obtained from the SASA values of our MD simulations. Alternatively, the loading capacity of CAuNPs of various sizes can also be theoretically estimated from their known geometric parameters, as detailed in Supporting Information Section S2. The theoretically estimated values agree well with the values obtained from MD simulations.

Although smaller CAuNPs provide sufficient surface area to accommodate the calculated DNA loading capacity (i.e., $N_{DNA}$), the instantaneous snapshots of the 1:2 complex show incomplete DNA attachment, with the terminal ends of DNAs remaining unbound from the CAuNP surface. For example, CAuNP_2nm has enough surface area for the attachment of two dsDNAs; however, both dsDNAs didn't exhibit complete attachment (see Fig. 1d). Similarly, despite CAuNP_3nm having the capacity to accommodate up to three dsDNAs, even two dsDNAs didn't completely attach to it (see Fig. 1e).

**Table 6.** The surface area of CAuNPs ($S_{CAuNP}$), single DNA coverage on the surface of CAuNPs ($S_{DNA}$), surface area of CAuNP occupied by repulsive zone ($S_{rep}$), total CAuNP's surface area occupied by DNA and the corresponding repulsive zone ($S_{DNA\_rep}$), and DNA Loading capacity per CAuNPs.

|  | $S_{CAuNP}$ (Å²) | $S_{DNA}$ (Å²) | $S_{rep}$ (Å²) | $S_{DNA\_rep}$ (Å²) | $N_{DNA}$ |
|---|---|---|---|---|---|
| **CAuNP_2nm** | 11642.34 | 3182.53 | 2546.02 | 5728.54 | ~2 (2.03) |
| **CAuNP_3nm** | 17890.12 | 3208.94 | 2567.15 | 5776.09 | ~3 (3.09) |
| **CAuNP_4nm** | 26012.84 | 3248.72 | 2598.97 | 5847.69 | ~4 (4.44) |

To understand the molecular origin of such behaviour, we measured the effective horizontal circumference of the CAuNPs for the attachment of the second dsDNA. Since a single dsDNA occupies more than half (i.e., > 50%) of the vertical circumference of the smaller CAuNPs, the remaining space along the vertical axis is inadequate for the attachment of a second dsDNA. As a result, the second dsDNA undergoes a twist from its initial orientation and wraps around the horizontal circumference of smaller CAuNPs. Here, the vertical attachment of the first dsDNA reduces the available horizontal circumference for the second dsDNA by occupying a some portion of the CAuNPs' horizontal circumference (see Fig 8). If the reduced horizontal circumference does not provide enough space for the complete attachment of the second dsDNA, incomplete binding would occur.

To check whether the above case occurs in our systems, we calculated the available horizontal circumference, by excluding the portion occupied by the first dsDNA with the repulsive zone, which is named as the effective horizontal circumference. The effective horizontal circumference is calculated using the following formula:

$$C_{h\_eff} = C_h - (W_{DNA} + D_{rep}) \quad (6)$$

Where, $C_h$, $C_{h\_eff}$, $W_{DNA}$, $D_{rep}$ are the original horizontal circumference of CAuNP, effective horizontal circumference of CAuNP, width of the DNA (i.e., horizontal cross section of DNA), and inter DNA repulsive distance.

The calculated $C_h$, $W_{DNA}$, $L_{DNA\_rep}$, and $C_{h\_eff}$ values are given in Table 7. As shown in table, the available horizontal circumference of both CAuNP_2nm and CAuNP_3nm is greater than the dsDNA length. This suggests that, for these smaller CAuNPs, the horizontal circumference is sufficient for the attachment of a second dsDNA. However, the second dsDNA is twisted but does not adopt a completely horizontal conformation. This may lead to a closer alignment of the second dsDNA (particularly at the ends) with the first, thereby increasing the repulsion between the two dsDNAs and causing the incomplete wrapping of both dsDNAs.

In the case of CAuNP_4nm, both dsDNAs attached in a vertical circumference and didn't exhibit any significant repulsion. As a result, this larger CAuNP may be able to accommodate the calculated loading capacity of dsDNAs. Further our results suggest that the orientation of DNA on CAuNPs plays a critical role in determining the loading capacity of DNA. Hence, the detailed MD simulation examining various orientations of multiple dsDNAs is required to validate this speculation, which are planned for future investigation.

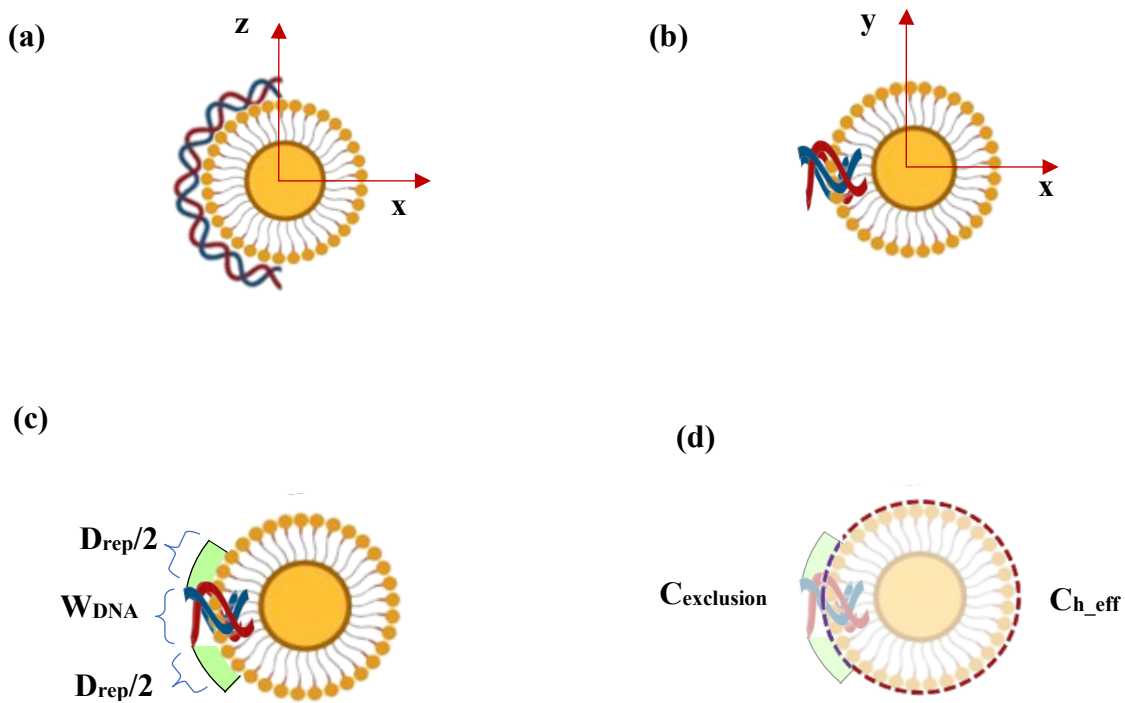

**Fig 8.** Schematic illustration of a single DNA attached along the vertical circumference of a CAuNP, shown in the normal projection onto (a) the XZ plane, and (b) XY plane. (c) the segment of the horizontal circumference occupied by vertically attached DNA, along with the associated repulsive region, (d) Illustration of the effective and excluded portions of the horizontal circumference of the CAuNP, where repulsive regions are highlighted in green, and the dotted violet and red lines represent the excluded and effective segments, respectively.

**Table 7.** Circumference of CAuNP ($C_h$), Width of the DNA ($W_{DNA}$), calculated effective horizontal circumference of CAuNP ($C_{h\_eff}$), and the length of the DNA along with its repulsive distance ($L_{DNA\_rep}$).

|  | $C_h$ (Å) | $W_{DNA}$ (Å) | $D_{rep}$ (Å) | $C_{h\_eff}$ (Å) | $L_{DNA\_rep}$ (Å) |
|---|---|---|---|---|---|
| **CAuNP_2nm** | 144.19 | 19.86 | 12 | 112.33 | 109.33 |
| **CAuNP_3nm** | 184.53 | 19.23 | 12 | 153.30 | 109.55 |

## 4) Conclusion

Overall, this study emphasizes the crucial role of AuNP size and CAuNP-to-DNA ratio in determining the wrapping nature of DNA around CAuNP and their binding affinity. Through all-atom MD simulations, we demonstrate that single dsDNA binding to the smaller CAuNP_2nm and CAuNP_3nm exhibits compact wrapping compared to larger CAuNP_4nm. However, when two dsDNAs interact with smaller CAuNPs, the mutual electrostatic repulsion between the free ends of dsDNA and the limited circumference of the CAuNPs lead to the twisting of dsDNAs from their initial orientation and the incomplete wrapping around the smaller CAuNPs. Also, the higher curvature of smaller CAuNP_2nm induces sharp bending in a single dsDNA, while smooth bending occurs when two dsDNAs interact with it. However, single and double dsDNAs bend smoothly around CAuNP_3nm and CAuNP_4nm. Further, our binding free energy calculation suggests that compared to smaller CAuNPs, the larger CAuNPs exhibit stronger binding affinity to DNAs due to their greater surface area and higher cationic charge density. Also, the loading capacity calculation shows that the maximum DNA loading capacity of CAuNPs increases with their size. Additionally, we developed a geometry-based theoretical method to estimate the DNA loading capacity of CAuNPs, which showed excellent agreement with the loading capacity calculated from the MD trajectory. Our study showed that smaller CAuNPs cannot reach their estimated loading capacity due to the vertical and horizontal orientations of DNA, which may lead to closer alignment of the two dsDNA molecules, and cause repulsion between them. This highlights the significance of DNA orientation in determining the loading capacity of CAuNPs.

In essence, our study demonstrates that smaller CAuNPs_2nm are optimal for compact wrapping of dsDNA, while larger CAuNP_4nm is suitable for holding multiple dsDNAs strongly. For gene therapy application requiring ultra-small CAuNPs of size around 2nm, a 1:1 CAuNP-to-DNA ratio would be the best choice to ensure the optimal dsDNA compaction. For enhanced dsDNA loading capacity, larger CAuNP_4nm with a 1:2 ratio or higher dsDNA per CAuNP would be the preferred strategy. Moreover, the DNA loading capacity formula provided in this paper will enable researchers involved in gene therapy to approximate the loading capacity of CAuNPs with varying sizes and ligand lengths.

**Supporting Information**

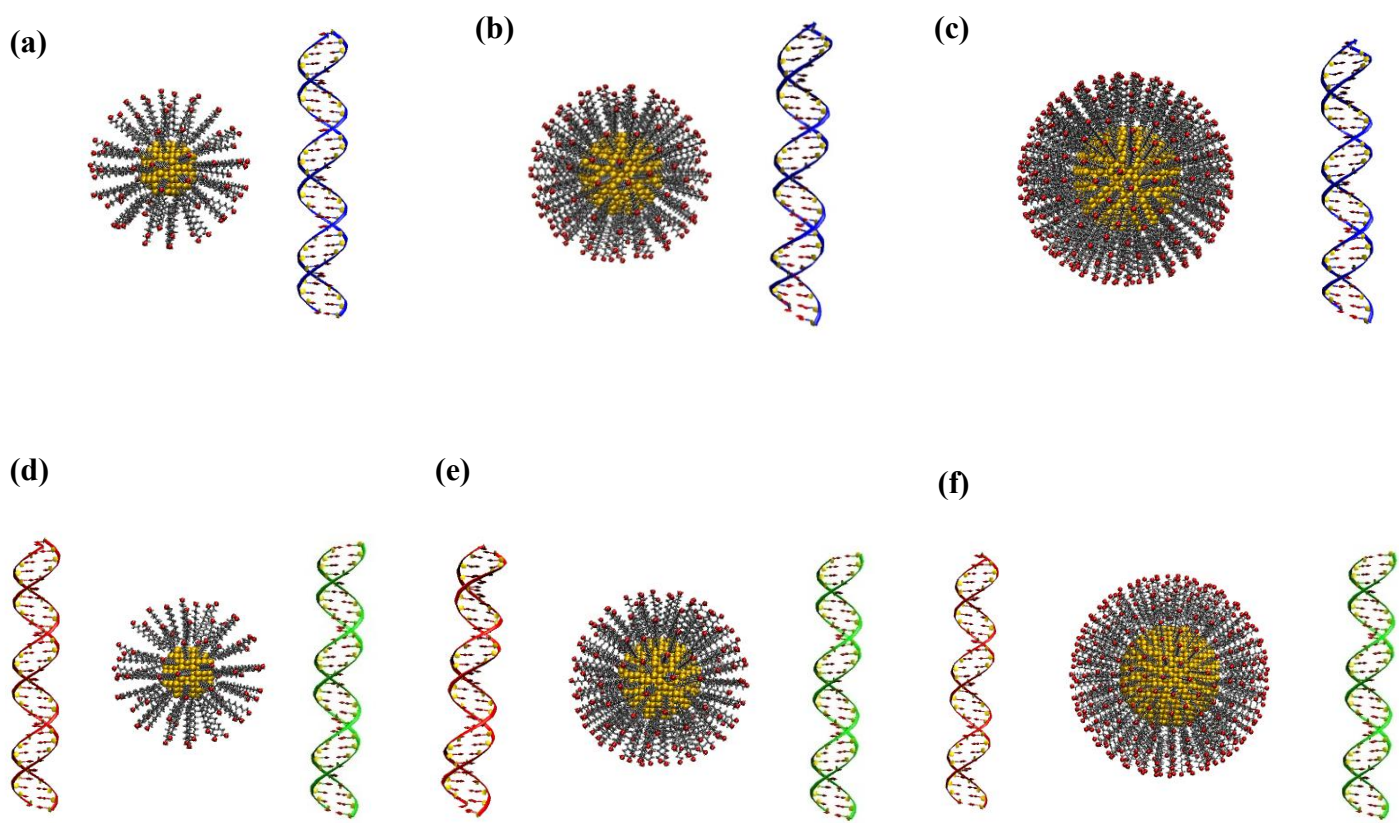

**Fig S1.** Snapshots of the initial configurations of (a) CAuNP_2nm, (b) CAuNP_3nm, and (c) CAuNP_4nm in the 1:1 complex, and (d) CAuNP_2nm, (e) CAuNP_3nm, and (f) CAuNP_4nm in the 1:2 complex. In a 1:2 complex, green-colored DNA is referred to as R-DNA (Right-side DNA), and red-colored DNA as L-DNA (left-side DNA).

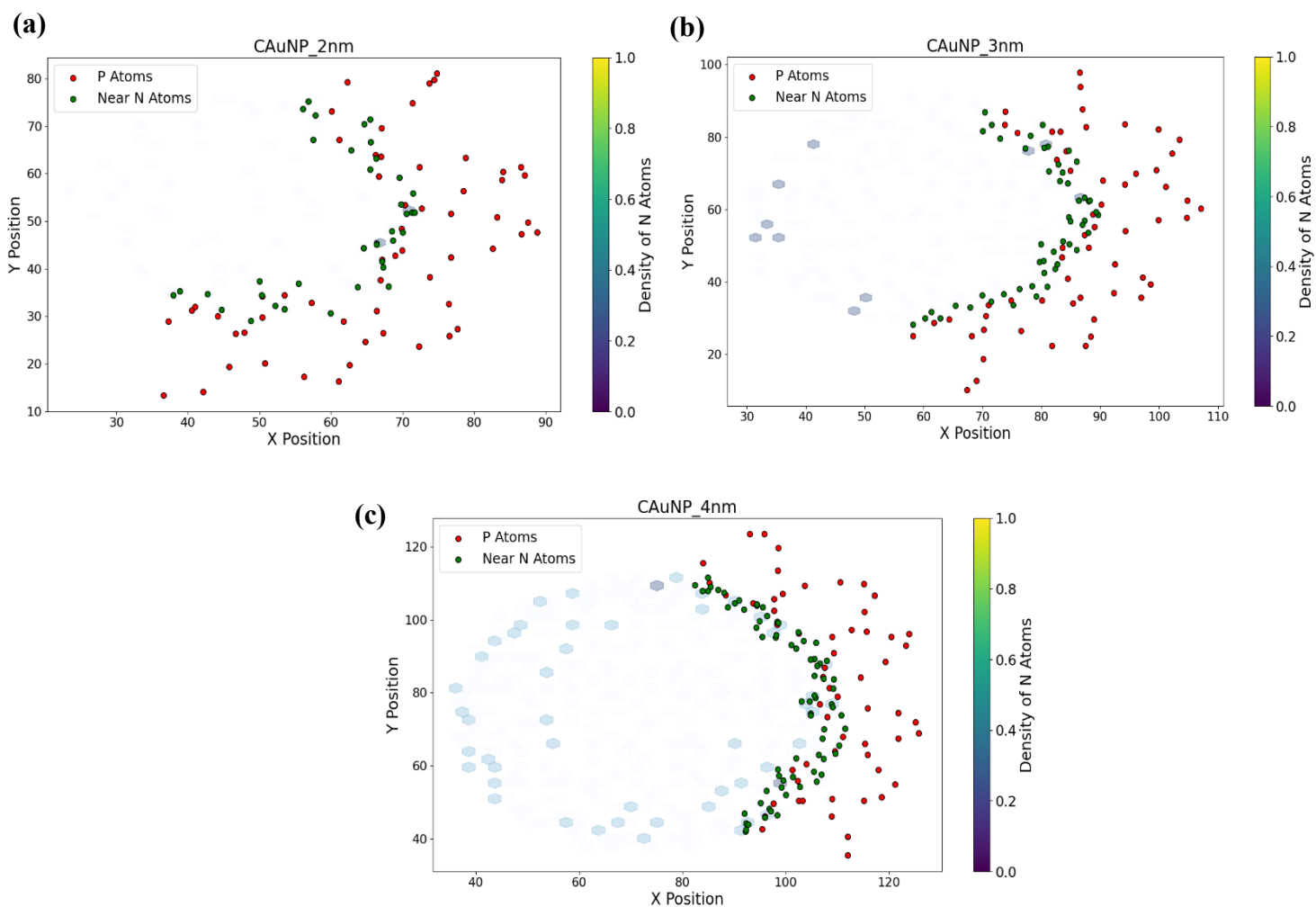

**Fig. S2.** Density plot illustrating the spatial distribution of P atoms in the backbone of DNA (red spheres) and N atoms in CAuNPs of different sizes: (a) CAuNP_2nm, (b) CAuNP_3nm, and (C) CAuNP_4nm. N atoms located within 5 Å of P atoms are highlighted as green spheres.

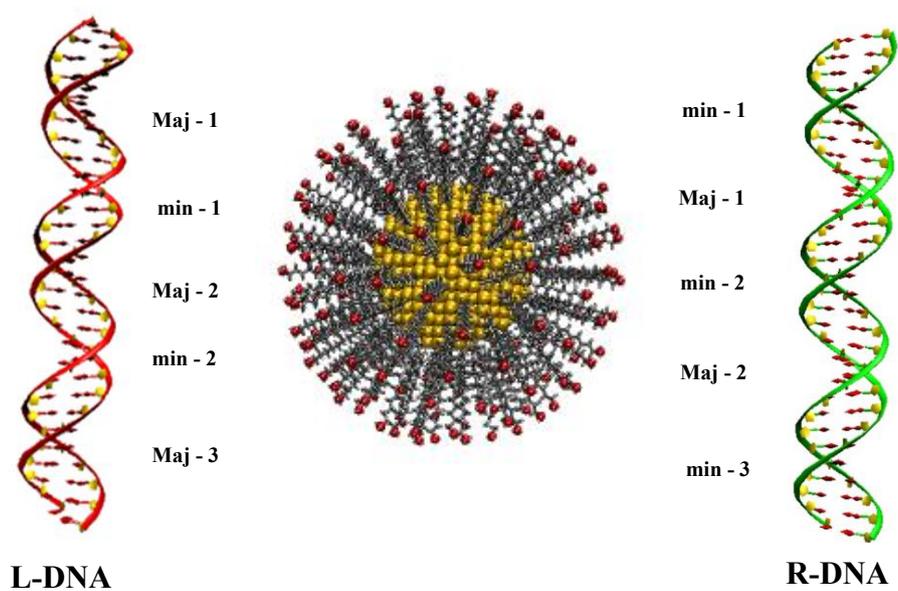

**Fig S3.** Structural model of the 1:2 DNA-CAuNP complex, illustrating the exposure of DNA grooves to CAuNP. Labels "Maj" and "Min" denote the major and minor grooves of the DNA, respectively.

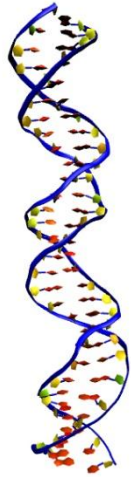

**Fig S4.** The snapshot of the bare DNA simulated in a water box with an appropriate number of ions i.e., the non-bend DNA. The water and ions were not shown here.

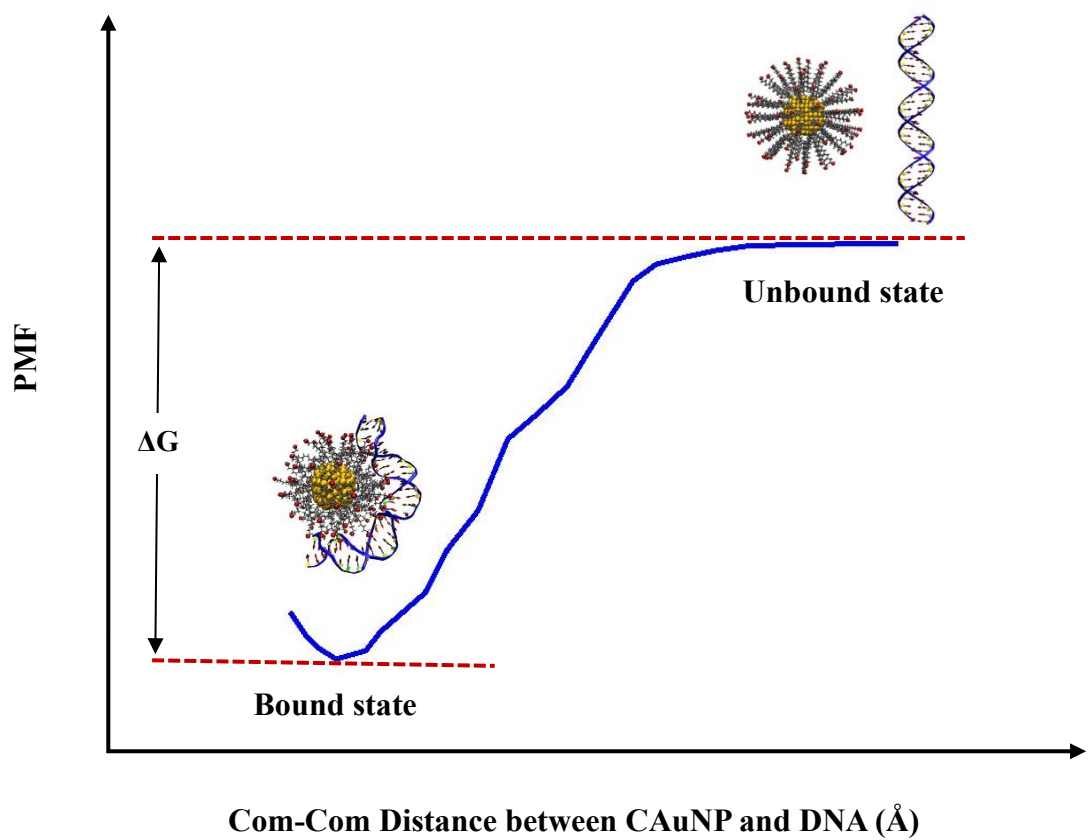

**Fig. S5.** Binding free energy is evaluated as the difference between the largest and smallest values of the PMF curve derived from the US simulations.

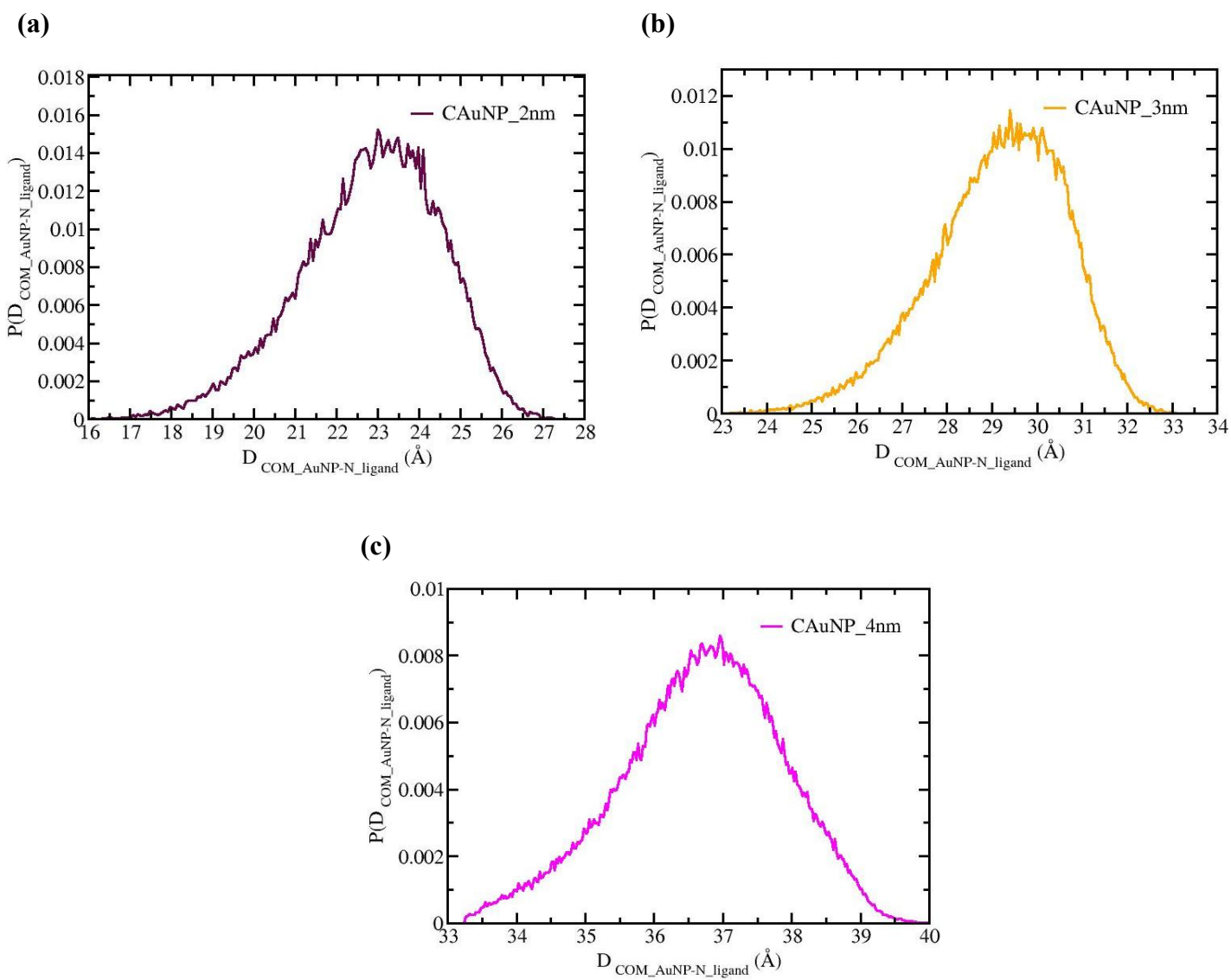

**Fig. S6.** Probability distribution of the center-of-mass distance between the AuNP and the nitrogen atoms of each cationic ligand for (a) CAuNP_2nm, (b) CAuNP_3nm, and (c) CAuNP_4nm.

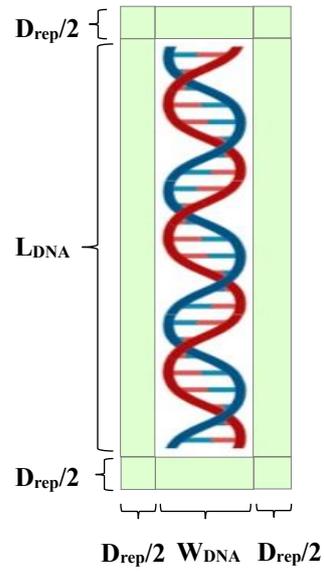

**Fig. S7.** Schematic representation of the repulsive region surrounding DNA on the surface of CAuNP. The repulsive area is considered as a rectangular region extending along the left and right transverse sides, as well as the top and bottom (longitudinal) ends of the DNA. The repulsive regions are highlighted in green.

Based on the **Fig. S7**, the length of the DNA along with its repulsive zone defined as follows:

$$= \frac{D_{rep}}{2} + L_{DNA} + \frac{D_{rep}}{2}$$

$$= L_{DNA} + D_{rep}$$

Similarly, the width of the DNA along with the repulsive zone is defined as:

$$= \frac{D_{rep}}{2} + W_{DNA} + \frac{D_{rep}}{2}$$

$$= W_{DNA} + D_{rep}$$

**Section S1. Step-by-Step Calculation of the Repulsive Zone Percentage (Eqs. 3–5)**

Using $L_{DNA}$ = 97 Å, $W_{DNA}$ = 20 Å, and $D_{rep}$ = 12 Å:

**Eq. 3. The area of DNA**

$$A_{DNA} = L_{DNA} \times W_{DNA} = 97 \times 20 = 1940 \text{ Å}^2$$

**Eq. 4. The area of DNA along with repulsive zone**

$$A_{DNA\_rep} = (L_{DNA} + D_{rep}) \times (W_{DNA} + D_{rep}) = (97 + 12) \times (20 + 12) = 3488 \text{ Å}^2$$

**Eq. 5. The area of repulsive zone alone**

$$A_{rep} = A_{DNA\_rep} - A_{DNA} = 3488 - 1940 = 1548 \text{ Å}^2$$

**Percentage of repulsive zone**

$$\% \text{ of Repulsive zone} = \frac{A_{rep}}{A_{DNA}} \times 100 = \frac{1548}{1940} \times 100 = 79.79 \% \ (\approx 80\%)$$

The percentage of the repulsive zone is therefore taken as approximately 80% for further calculations.

**Section S2. Theoretical calculation of loading capacity**

In this approach, the surface areas of CAuNP and DNA are calculated from their known geometric parameters and used to theoretically estimate the DNA loading capacity of CAuNPs of various sizes.

The surface area of the CAuNP can be calculated using the following equation:

$$S_{T\_CAuNP} = 4\pi r^2_{T\_CAuNP} \tag{S1}$$

Where, $S_{T\_CAuNP}$, and $r_{T\_CAuNP}$ are the theoretically calculated surface area of CAuNP, and the radius of the CAuNP, respectively.

The radius of a functionalized CAuNP can be determined by summing of three main contributions: (i) the radius of bare AuNP, (ii) the distance between the Au surface atom and the base atom (sulphur) of the alkanethiol, which is anchored to the CAuNP surface through a covalent Au–S bond, and (iii) the alkanethiol length.

While calculating the length of the alkanethiol, it is important to account for the reduction in ligand length due to tilting during self-assembly on the CAuNP surface. Several studies have reported the tilt angle of alkanethiols with varying chain lengths and terminal groups attached to AuNPs of different sizes [38,47]. The tilt angle is defined as the angle between the alkane chain axis and the surface normal of the attached surface. It can be used to determine the vertical length of the alkanethiol after tilting, using the formula below.

$$l_{T\_ligand} = l_{ligand} \times cos\theta \tag{S2}$$

Where, $l_{T\_ligand}$, $l_{ligand}$ and $\theta$ are the vertical length of the ligand after tilting, original length of the ligand and the tilt angle respectively.

The surface area of DNA, along with the repulsive zone, is determined using the below formula

$$S_{T\_DNA} = (L_{DNA} + D_{Rep}) \times (W_{DNA} + D_{Rep}) \tag{S3}$$

Where, $L_{DNA}$, $W_{DNA}$, and $D_{rep}$ are the length, width of the DNA, and the inter DNA repulsive distance, respectively.

The basis of Equation S3 is illustrated in detail in Fig. S7. The width of DNA is constant around 20 Å, and the length of the DNA can be estimated by multiplying the number of base pairs of the DNA with the distance between the adjacent base pairs, which is the constant value of 3.4 Å. The repulsive zone surrounding DNA is ranges from 10 to 15 Å.

To validate this theoretical framework, the surface areas of the CAuNP and DNA considered in this study were calculated from their known geometric parameters using the formulas presented in Equations S1, and S3. The length of the alkanethiol used in this study and the distance between the Au and the sulphur atoms of the alkanethiol are 15.7 Å and 2.3 Å, respectively. Also, the tilt angle of alkanethiol with similar chain lengths has been reported as ~ 30° upon self-assembly on AuNPs[47]. Hence, the vertical length of alkane thiol after tilting was calculated to be 13.59 Å using Equation S2. For a 30 base-pair DNA, the theoretical length is estimated as 102 Å (30 X 3.4 Å). The calculated $r_{CAuNP}$, $S_{T\_AuNP}$ and $S_{T\_DNA}$ values were shown in Table S1.

Additionally, to ensure the accuracy of Equation S2 and the reported tilt angle in determining the vertical length of the alkanethiol after tilting, we measured the CAuNP radius before and after alkanethiol self-assembly on the CAuNP by calculating the average distance between the center of mass of the AuNP and the nitrogen atom of the cationic ligand. The difference in radius before and after self-assembly directly quantifies the reduction in alkanethiol length caused by tilting. This reduction value is then subtracted from the original length of the alkanethiol to determine its vertical length after titling during self-assembly (see Table S2). We found that the alkanethiol length derived from the reduction in radius values is nearly equal to the value estimated based on Equation.

Also, from the Table S1, it is apparent that the theoretically estimated value matches the loading capacity obtained from our MD simulation. Therefore, this theoretical approach can be used to determine the loading capacity of CAuNPs of various sizes and the alkanethiol of various lengths with the cationic end group.

**Table S1.** Radius obtained from MD simulation and theoretically calculated radius ($r_{T\_CAuNP}$) of CAuNPs of various sizes, surface area of CAuNP ($S_{T\_CAuNP}$), single DNA coverage on CAuNP along with the repulsive zone ($S_{T\_DNA}$), and DNA Loading capacity per CAuNPs.

|  | r (Å) | $r_{T\_CAuNP}$ (Å) | $S_{T\_CAuNP}$ (Å²) | $S_{T\_DNA}$ (Å²) | DNA Loading capacity per CAuNPs |
|---|---|---|---|---|---|
| CAuNP_2nm | 22.95 | 25.89 | 8423.13 | 3721.25 | ~2 (2.26) |
| CAuNP_3nm | 29.37 | 30.89 | 11990.73 | 3721.25 | ~3 (3.22) |
| CAuNP_4nm | 36.95 | 35.89 | 16186.64 | 3721.25 | ~4 (4.34) |

**Table S2.** The radius of CAuNPs before and after self-assembly, the reduction in the ligand length due to self-assembly, length of the ligand after titling and radius of CAuNP.

|  | Radius of CAuNPs (Å) | | Reduction in ligand length due to self-assembly (Å) | Length of ligand after titling (Å) | Radius of the CAuNP (Å) |
| --- | --- | --- | --- | --- | --- |
|  | Before self-assembly | After self-assembly | | | |
| CAuNP_2nm | 26.12 | 23.25 | 2.87 | 12.83 | 25.13 |
| CAuNP_3nm | 31.96 | 29.58 | 2.38 | 13.32 | 30.41 |
| CAuNP_4nm | 37.98 | 35.96 | 2.02 | 13.68 | 35.98 |